\documentstyle[12pt]{article}
\begin{document}

\begin{center}
\large{\bf{Study of deuteron-proton charge exchange reaction 
\\ at small transfer momentum}}
\\
\vspace{1cm}
{N.B.Ladygina 
$^{1}$
A.V.Shebeko$^{2}$
}
\end{center}

\noindent{\small\it
$^1$ Laboratory of High Energies,
Joint Institute for Nuclear Research, 141980 Dubna, Russia,
\\
$^2$NSC Kharkov Institute of Physics \& Technology, 61108 Kharkov, Ukraine.}

\sloppy

\begin{abstract}
The charge exchange reaction $pd\to npp$ at 1 GeV projectile proton
energy is studied in the multiple-scattering expansion technique.
This reaction is considered in a special kinematics, when the transfer
momentum from the beam proton to fast neutron is close to zero. 
The differential
cross section and a set of polarization observables are calculated.
It was shown that contribution of the final state interaction between
two protons is very significant.
\end{abstract}

\section{Introduction}

Within the several last decades the deuteron- proton charge exchange
reaction has been studied both from  experimental and theoretical
point of view. A considerable interest to this reaction is connected,
first of all, with an opportunity to extract some information about
spin-dependent part of the elementary nucleon-nucleon charge
exchange amplitudes. This idea was suggested by Pomeranchuk \cite {Pom}
already in 1951, but up to now it is actual. Later this supposition
has been developed in \cite {Dean, Wil, Car}. It was shown, that in the
plane-wave impulse approximation (PWIA) the differential cross
section and tensor analyzing power $T_{20}$ in the dp-charge exchange
reaction are actually fully determined  by the spin-dependent
part of the elementary $np\to pn$
 amplitudes.
 The analogous 
result was obtained in \cite {Kaptari}, where this process has been
studied in the Bethe-Salpeter formalism.

The differential cross section of the $dp\to npp$ reaction at 3.34 GeV/c
deuteron beam has been measured in 1970's in the 1m hydrogen bubble
chamber of the JINR Synchrophasatron \cite {Gl, Al}. However,
the obtained statistics is not sufficient to evaluate the magnitude of the
spin-dependent part of the elementary amplitude.
Nowadays the experiment on the study of the dp-charge exchange reaction at
the small transfer momentum in GeV-region is planned at ANKE setup
 at COSY \cite {cosy}. The aim of this experiment is to provide the
information about the spin-dependent np-elastic scattering
amplitudes in the energy
region, where the phase-shift analysis data are absent. 

From our point
of view, under kinematical conditions proposed in this experiment,
when momentum of the emitted neutron has the same direction and magnitude
as the beam proton (in the deuteron rest frame), and relative momentum
of  two protons is very small, the final state interaction (FSI)
effects play very important role. The contribution of the D-wave
in the DWF into differential cross section in this kinematics
must be negligible \cite {vvg}. However, for the polarization observables
the influence of the D-component can be significant.

The goal of our paper is to study the importance of the D-wave and
FSI effects under kinematical conditions of the planned experiment.
We consider $pd\to npp$ reaction in the approach, which has been used
by us to describe the pd breakup process at 1 GeV projectile proton
energy \cite {LSh}. This approach is based on the Alt-Grassberger-Sandhas
 formulation
of  the multiple-scattering theory for the three-nucleon system.
The matrix inversion method \cite {HaTa, BJ}
has been applied to take account of the FSI contributions. Since
unpolarized and polarized mode of the deuteron beam are supposed to be
employed 
in the experiment, we also calculate both the differential cross
section and a set of the polarization observables.
It should be noted, in this paper we have not considered the Coulomb
interaction in the (pp)-pair. This problem is nontrivial and
requires a special investigation.

The paper is organized as following. In sect.2 the short description
of the general theoretical formalism is given. The special kinematics,
when the transfer momentum from the beam proton to neutron is close to zero,
is considered in sect.3. The results of our calculations for the
differential cross section and  polarization observables
are presented in sect.4. The figures in this section demonstrate
the behaviour of this observables obtained in the PWIA and PWIA+FSI
 with D-wave in the DWF included and without it. The significant
dependence of the calculation results on the elementary NN-amplitudes
is also shown. We conclude with sect.5.

\section{Theoretical Formalism}

 In accordance to the three-body collision theory
 let us write the matrix element of the deuteron proton
 charge exchange reaction
\begin{equation}
\label{r}
p(\vec p)+d(\vec 0)\to n(\vec p_1)+p(\vec p_2)+p(\vec p_3)
\end{equation}
in the following form \cite {LSh}
\begin{eqnarray}
\label{am}
U_{pd \to npp}&=&\sqrt {2} <123|[1-(2,3)][1+t_{23}(E-E_1) 
g_{23} (E-E_1)]t_{12}^{sym}|1(23)>,
\end{eqnarray}
where the operator $g_{23} (E-E_1)$ is a free propagator for the
(23)-subsystem and the scattering operator $t_{23}(E-E_1)$
satisfies the Lippmann-Schwinger (LS)
equation with two-body force operator $V_{23}$ as  driving term
\begin{eqnarray}
\label{LS}
t_{23}(E-E_1) = V_{23} + V_{23} g_{23}(E-E_1) t_{23}(E-E_1) .
\end{eqnarray}
Here $E$ is the total energy of the three-nucleon system 
$E=E_1+E_2+E_3$.

Let us rewrite the matrix element (\ref{am}) indicating explicitly
the particle quantum numbers,
\begin{eqnarray}
U_{pd\to npp}=\sqrt {2}
<\vec {p_1} m_1 \tau_1,\vec {p_2} m_2 \tau_2,\vec {p_3} m_3 \tau_3|
[1-(2,3)] \omega_{23} t^{sym}_{12} |\vec {p} m \tau ,\psi _{1 M_d 0 0} (23)>,
\nonumber
\end{eqnarray}
where $\omega_{23}=[1+t_{23}(E-E_1) g_{23} (E-E_1)]$ and
the the spin and isospin  projections denoted as
$m$ and $\tau$, respectively. The permutation operator for two nucleons
$(i,j)$ was introduced here. The operator $t_{12}^{sym}$ is symmetrized
NN-operator, $t_{12}^{sym}=[1-(1,2)]t_{12}$.
Inserting the unity 
\begin{eqnarray}
{\bf 1}=\int d\vec p^\prime
|\vec p^\prime m^\prime\tau ^\prime>
<\vec p^\prime m^\prime\tau ^\prime|,
\nonumber
\end{eqnarray}
we get  the following expression for the reaction amplitude
\begin{eqnarray}
\label {1}
{\cal J}&=&(-1)^{1/2+\tau_3^\prime }
<{1\over 2} \tau_2 {1\over 2} \tau_3|T \tau_2+\tau_3>
<{1\over 2} \tau_2^\prime {1\over 2} \tau_3^\prime |T \tau_2+\tau_3>
<{1\over 2} \tau_1 {1\over 2} \tau_2^\prime |T^\prime M_{T^\prime }>
\nonumber\\
&&<{1\over 2} \tau {1\over 2} -\tau_3^\prime |T^\prime M_{T^\prime }>
<{1\over 2} m_2 {1\over 2} m_3|S M_S>
 <{1\over 2} m^\prime _2 {1\over 2} m^\prime _3|S M^\prime _S>
\nonumber\\
&&\int d\vec p _0 {^\prime }
\left<\vec p_0, S M_S\left|1 + m_N \frac{t^{ST}(E-E_1)}
{\vec {p_0} ^2 -\vec {p_0}^{\prime 2}+i0 }\right|
\vec p _0 {^\prime }, S M^\prime _S \right>
\nonumber\\
&&<\vec {p_1} m_1,(\vec {p_0}^\prime + \vec q/2)~ m^\prime _2|
t^{T^\prime}_{sym}(E - E_3^\prime)
|\vec p m, (\vec {p_0}^\prime -\vec q/2)~ m^{\prime \prime}>
\\
&&<m^{\prime\prime} m^\prime_3|\psi _{1M_d}(\vec p_0 {^\prime } -\vec q/2)>
~-~
(2\leftrightarrow 3),
\nonumber
\end{eqnarray}
where $E_3^\prime=\sqrt{m_N^2+(\vec p_0{^\prime} -\vec q/2)^2}$,
$m_N$ the nucleon mass , and  we have 
introduced the  momentum transfer
$\vec q=\vec p -\vec p_1$,  relative momenta
$\vec p_0={1\over 2}(\vec p_2 -\vec p_3)$ and
$\vec {p_0}^\prime ={1\over 2}(\vec {p_2}^\prime  -\vec {p_3}^\prime)$.
Henceforth, all summations over dummy discrete indices are implied.

In momentum representation the DWF  $\psi _{1 M_d}(\vec k)$  with
 spin projection $M_d$ is written as
\begin{eqnarray}
|\psi _{1 M_d}(\vec k)>=\sum _{L=0,2}\sum_{M_L=-L}^{L}
<L M_L 1 M_s|1 M_d> u_L (k) Y_L^{M_L}(\hat k) | 1 M_s>,
\end{eqnarray}
with the spherical harmonics $Y_L^{M_L}(\hat k)$ and the Clebsh-Gordon
coefficients in standard form.
In our calculations, we have employed the following parameterizations of
the S- and D- state wave functions
\begin{eqnarray}
u_0 (p)=\sqrt{{2\over\pi}}\sum _{i}\frac {C_i}{\alpha_i ^2 +p^2}~~,~~
u_2 (p)=\sqrt{{2\over\pi}}\sum _{i}\frac {D_i}{\beta_i ^2 +p^2}
\end{eqnarray}
as proposed in refs. \cite {Par,B,CD} .

We assume that $\tau = \tau_2 =\tau_3 = 1/2 $ and
$\tau_1 = - 1/2 $. Then the isotopic coefficient can be calculated
and Eq.(\ref{1}) is simplified
\begin{eqnarray}
\label {11}
{\cal J}&=&{1\over 2}
<{1\over 2} m_2 {1\over 2} m_3|S M_S>
 <{1\over 2} m^\prime _2 {1\over 2} m^\prime _3|S M^\prime _S>
\nonumber\\
&&<L M_L 1 {\cal M_S} |1 M_d>
 <{1\over 2} m^{\prime\prime }  {1\over 2} m^\prime _3|1 {\cal M _S}>
\\
&&\int d\vec p _0 {^\prime }
<\psi^{(-)}_{\vec p_0 S M_S T M_T}|\vec p_0 {^\prime } S M^\prime _S T M_T>
u_L (|\vec p_0^\prime -\vec q /2|)Y_L^{M_L}
(\widehat {\vec p_0^\prime -\vec q /2})
\nonumber\\
&&<\vec {p_1} m_1,(\vec {p_0}^\prime + \vec q/2)~ m^\prime _2|
t^0_{sym}(E - E_3^\prime)-t^1_{sym}(E - E_3^\prime)
|\vec p m, (\vec {p_0}^\prime -\vec q/2)~ m^{\prime \prime}>
\nonumber\\
&&~-~
(2\leftrightarrow 3),
\nonumber
\end{eqnarray}

The wave function of the final p-p pair
\begin{eqnarray}
<\psi^{(-)}_{\vec p_0 S M_S T M_T}|\vec p _0 {^\prime } S M^\prime _S T M_T>
=
\delta (\vec {p_0} - \vec {p_0}^\prime)\delta _{M_s M_s^\prime}+
\nonumber\\
\frac{m_N}{\vec {p_0} ^2 -\vec {p_0}^{\prime 2}+i0 }
< \vec p_0 S M_S | t^{ST}| \vec p _0 {^\prime } S M^\prime _S >
\end{eqnarray}
contains the FSI part, which can be taken  in different ways.

In this paper we use the matrix inversion method (MIM) suggested in refs.
\cite {HaTa, BJ} and
applied to study the deuteron  electro-disintegration \cite {Sch1,Sch2}
and deuteron proton breakup process \cite {LSh}.
As in ref. \cite {Sch1}, we consider the truncated partial-wave expansion,
\begin{eqnarray}
\label {psi}
&&<\psi^{(-)}_{\vec p_0 S M_S T M_T}|\vec p_0 {^\prime } S M^\prime _S T M_T>=
\delta _{M_S M^\prime _S} \delta (\vec p_0 -\vec p_0 {^\prime})+
\\
&&\sum _{J=0}^{J_{max}} \sum _{M_J=-J}^{J}
Y_l^\mu (\hat p_0) <l \mu S M_S|J M_J>
\psi _{l l^\prime} ^\alpha (p_0 {^\prime})
<l^\prime \mu ^\prime S M_S ^\prime|J M_J>
{Y} _{l^\prime}^{\ast \mu ^\prime} (\hat p_0{^\prime}),
\nonumber
\end{eqnarray}
where $J_{max}$ is the maximum value of the total angular momentum in
p-p partial waves and $\alpha =\{ J,S,T\}$ is the set of conserved quantum
numbers.
The radial  functions $\psi^\alpha _{ll^\prime}(p_0{^\prime})$ are related via
\begin{eqnarray}
\label {2}
\psi _{l l^\prime} ^\alpha (p_0 {^\prime})=\sum _{l^{\prime\prime}}
O_{l l^{\prime\prime}}\varphi _{l^{\prime\prime}l^\prime}^\alpha
(p_0^{\prime})-
\frac {\delta (p_0 ^{\prime} -p_0)}{p_0^2} \delta _{l l^\prime}
\end{eqnarray}
to the partial wave functions
$\varphi_{l^{\prime\prime}l^\prime}^\alpha (p_0^\prime)$, which
have the asymptotics of  standing waves. The coefficients
$O_{ll^{\prime\prime}}$ can be expressed in  terms of the
corresponding phase shifts and mixing parameters \cite {Sch1}.

Within the MIM, the functions $\varphi_{ll^\prime}^\alpha (p_0^\prime)$
can be represented as
\begin{eqnarray}
\varphi _{ll^\prime} ^\alpha (p_0 {^\prime})=\sum _{j=1}^{N+1}
B_{ll^\prime} ^\alpha (j) \frac {\delta (p_0 ^{\prime} -p_j)}{p_j^2},
\end{eqnarray}
where the coefficients $B_{ll^\prime} ^\alpha (j)$ fulfill a  set of
 linear algebraic equations approximately equivalent to the
$LS$ integral equation for the $p-p$ scattering problem
\footnote {We neglect here the Coulomb interaction in the (pp)-pair.}.
Here $N$ is the dimension of this set,
 $p_j$ are the grid points associated with the Gaussian nodes over the
interval [-1,1] and $p_{N+1}=p_0$
(details can be found in ref. \cite {KFTI}).
It should be noted that in this way the nucleon wave function is 
expressed by a series of
 $\delta$-functions allowing one to reduce a triple integral
 in Eq. (\ref{11}) to a double one. In addition, the method offers the opportunity to
 consider the nucleon wave function in the continuum directly in momentum
space what simplifies all subsequent calculations.

\section{Collinear geometry}

In this paper we consider the special kinematics, when
transfer momentum  $\vec q=\vec p -\vec p_1 $ is close to zero.
In other words, the neutron momentum has the same value and
direction as the beam proton. In fact, since the difference between
proton and neutron masses and deuteron binding energy take place,
the transfer momentum is not  exactly zero, $q\approx 1.8$ MeV/c. But 
because of this value is very small and has no significant influence
on the results, we shall suppose $q=0$ in the subsequent calculations.
                 
Under such kinematical conditions one can neglect the dependence
of the high-energy nucleon-nucleon  matrix $t_{NN}(E-E_3^\prime)$ in
Eq.(\ref {11}) on the internal nucleon-nucleon momentum in the deuteron
and express it in the center-of-mass
system (c.m.s.) through
 three independent amplitudes 
\begin{eqnarray}
t_{NN}^{cm}(\vec q =0)=A +
(F-B) (\vec\sigma_1 \hat Q^*) (\vec\sigma_2 \hat Q^*)+
B  (\vec\sigma_1 \vec\sigma_2),
\end{eqnarray}
where 
\begin{eqnarray}
\hat Q^*=\frac {{\vec p} ^* +\vec {p^\prime} ^*}{|{\vec p} ^* +\vec {p^\prime} ^*|}=
\hat p^*
\nonumber
\end{eqnarray}
with the fast proton (neutron) momentum $\vec p ^*$. 
We use some results of relativistic potential theory \cite {Hell76, Garc77}
to relate this NN t-matrix in the c.m.s. with that in the frame of interest
(see, also \cite {LSh}).
\begin{eqnarray}
\label{33}
<m_1 m_2^\prime ,\vec p \vec p_0^\prime|t|
\vec p\vec p_0^\prime,m m^{\prime\prime }>&=&
NN^\prime F
<m_1|D^\dagger (\vec u,\vec p)|\mu_1>
<m_2^\prime|D^\dagger (\vec u,\vec p_0^\prime)|\mu_2^\prime>
\\
&&<\mu_1\mu_2^\prime |t_{cm}(\vec p^*)|\mu\mu^{\prime\prime}>
<\mu|D (\vec u,\vec p)|m>
<\mu^{\prime\prime }|D(\vec u,\vec p_0^\prime)|m ^{\prime\prime}>,
\nonumber
\end{eqnarray}
where $D$ is a Wigner rotation operator in the spin space and
$u$ is a four-velocity. In our kinematical situation, when
$\vec p=\vec p_1 \gg \vec p_0^\prime $, each of these operators
is slightly different from the unit operator, so that in a good
approximation  the $t_{NN}$-matrix in the frame of interest has the same
 spin structure. The product of
the normalization factors $N$ and $N^\prime $ and kinematical
factor $F$ is 
\begin{eqnarray}
NN^\prime F=\frac{m_N+E_p}{4E_p},
\end{eqnarray}
In such a way, we have following relation for high-energy NN t-matrix
in the different frames of reference
\begin{eqnarray}
<m_1m_2^\prime|t(\vec p,\vec p_0^\prime)|m m^{\prime\prime}>
=\frac{m_N+E_p}{4E_p}
<m_1m_2^\prime|t_{cm}(\vec p^*)|m m^{\prime\prime}>
\end{eqnarray}

To evaluate such quantities without their momentum angular decomposition
 we use the 
phenomenological model suggested by Love and Franey in refs. \cite {LF}.
In this approach the corresponding matrix elements are expressed through the
effective NN-interaction operators sandwiched between the initial and
final plane-wave states, that enables us to extend this construction to
the off-shell case. Obviously, such  off-shell extrapolation does not
change the general spin structure.

Since the (pp)-pair belongs to the isotriplet, one can anticipate
that the FSI in the $^1S_0$ state is prevalent at comparatively small
$p_0$-values.
In such a way we get the following expression for amplitude of
the dp charge exchange process
\begin{eqnarray}
\label{ampl}
{\cal J}&=&{\cal J}_{PWIA}+{\cal J }_{^1S_0}
\nonumber\\
\nonumber\\
{\cal J}_{PWIA}&=&\frac {m_N+E_p}{4E_p} <L M_L 1 {\cal M_S}|1 M_D>
u_L ( p_0^\prime )
Y_L^{M_L}(\widehat { p_0^\prime })
\nonumber\\
&&\Bigl\{ <{1\over 2} m_2^\prime {1\over 2} m_3|1 {\cal M_S}>
< m_1 m_2|
t_{cm}^0 (\vec p^*) -t_{cm}^1(\vec p^*)
| m m_2 ^ \prime >-
\nonumber\\
&&<{1\over 2} m_2^\prime {1\over 2} m_2|1 {\cal M_S}>
< m_1 m_3|
t_{cm}^0(\vec p^*) -t_{cm}^1(\vec p^*)
| m m_2 ^ \prime > \}
\\
\nonumber\\
{\cal J}_{^1S_0}&=&\frac {(-1)^{1-m_2 -m_2^\prime}}{4\pi }
\frac {m_N+E_p}{4E_p} <L M_L 1 {\cal M_S}|1 M_D>
\delta _{m_2 ~ -m_3}
<{1\over 2} m^{\prime\prime } {1\over 2} -m_2^\prime|1 {\cal M_S}>
\nonumber\\
&&\int dp _0 {^\prime } p _0 {^\prime }	^2
\int d \hat p_0 {^\prime } 
u_L ( p_0^\prime )Y_L^{M_L}(\widehat {p_0^\prime })
\psi _{00} ^{001} (p_0^\prime )
\nonumber\\
&&< m_1 m _2^\prime |
t_{cm}^0(\vec p^*) -t_{cm}^1(\vec p^*)
| m m ^ {\prime\prime }>
\nonumber
\end{eqnarray}

One can see, we have very simple integral over angular variables
$\hat p_0^\prime$. As the result of this integration we get following
relation for ${\cal J}_{^1S_0}$
\begin{eqnarray}
\label{ampl2}
{\cal J}_{^1S_0}&=&\frac {(-1)^{1-m_2 -m_2^\prime}}{\sqrt {4\pi }}
\frac {m_N+E_p}{4E_p} \delta _{m_2 ~ -m_3}
<{1\over 2} m^{\prime\prime } {1\over 2} -m_2^\prime|1 M_D> 
\nonumber\\
&&< m_1 m _2^\prime |
t_{cm}^0(\vec p^*) -t_{cm}^1(\vec p^*)
| m m ^ {\prime\prime } >
\int dp _0 {^\prime } p _0 {^\prime }	^2
\psi _{00} ^{001} (p_0^\prime ) u_0(p_0^\prime)
\end{eqnarray}
Note, the integral over radial variable $p_0^\prime$ also has no any
difficulties, since  $\psi _{00}^{001}(p_0^\prime)$ contains 
$\delta$-functions.

\section{Results and discussions}

We define unpolarized  $2\to3$  cross section by the standard manner
\begin{eqnarray}
\sigma (dp\to npp)=(2\pi)^4 {E_p\over p} \int d\vec {p_1}d\vec {p_2}
d\vec {p_3}|\overline {\cal J}|^2 \delta^4(4-momentum),
\end{eqnarray}
where $|\overline {\cal J}|^2=1/6 Tr(\cal J\cal J ^+)$ is the
square of the process amplitude averaged over all particles spin
states. Using the $\delta ^3$-function to eliminate the $\vec p_3$
integration and changing variables from  $\vec p_1$ to $\vec q$ we 
have
\begin{eqnarray}
\sigma (dp\to npp)=(2\pi)^4 {E_p\over p} \int d\vec {q} d\vec {p_2}
|\overline {\cal J}|^2 
\delta (m_d+E_p-\sqrt {m_N^2+(\vec p -\vec q )^2}-E_2-
\sqrt {m_N^2+(\vec q -\vec p_2)^2}).
\end{eqnarray}
Taking $p_2, q \ll p$ this expression can be reduced to
\begin{eqnarray}
\label{sigma}
\sigma (dp\to npp)=(2\pi)^6 {{E_p^2}\over {2p^2}} \int d{q^2}d{p_2}
d cos \theta_2  p_2^2
|\overline {\cal J}|^2 
\end{eqnarray}
We define general spin observable related with polarization of 
initial particles in terms of the Pauli $2 \times 2$ spin matrices $\sigma$
for the proton and a set of spin operators $S$ for deuteron \cite {Ohls}
as following
\begin{eqnarray}
C_{\alpha\beta}=\frac {Tr ({\cal J}\sigma _\alpha S_\beta {\cal J})}
{Tr ({\cal J} {\cal J}^+) },
\end{eqnarray}
where indices $\alpha$ and $\beta$ refer to the proton and deuteron
polarization, respectively; $\sigma _0$ and $S_0$ corresponding to the
unpolarized particles are the unit matrices of two and three dimensions.
In such a way, Eqs.(\ref {ampl}- \ref {ampl2}) for dp- charge exchange
amplitude  enables us  to get the relation for any variable of this
process taking into account two slow protons final state interaction
in $^1S_0$ -state. So, we have following expression for the spin- averaged
squared amplitude
\begin{eqnarray}
\label{c0}
C_0&\equiv & Tr({\cal J}{\cal J}^+)={1\over {4 \pi}}\left( \frac{m_N+E_p}{2E_p}
\right) ^2\{ 2(2B^2+F^2)({\cal U}^2(p_2)+w^2(p_2))+
\\
&&(F^2-B^2)w(p_2)(w(p_2)-2\sqrt 2 Re{\cal U}(p_2))(3cos^2\theta_2-1)\},
\nonumber
\end{eqnarray}
where ${\cal U}(p_2)=u(p_2)+\int dp_0^\prime {p_0^\prime }^2
\psi _{00}^{001}(p_0^\prime ) u(p_0^\prime )$
is the S-component of the DWF  corrected on the FSI of the (pp)-pair.
One can see, the $C_0$ is independent on the neutron and proton
azimuthal angles $\phi _q$ and $\phi_2$. To obtain Eq.(\ref {sigma}) 
we have considered this fact and performed the integration over azimuthal 
angles. 
We use a right-hand coordinate system defined in accordance to the 
Madison convention \cite {mad}. The quantization $z$-axis is along
the beam proton momentum $\vec p$. Since the direction of 
$\vec p \times \vec p_1$ is undefined in the collinear geometry, 
we choose the $y$-axis normal to  the beam momentum. Then
third axis is $\vec x =\vec y\times \vec z$.

The tensor analyzing power can be presented in the following form 
\begin{eqnarray}
C_{0,yy}\cdot C_0&=&{1\over 2}{1\over {4 \pi}}\left( \frac{m_N+E_p}{2E_p}
\right) ^2
\{4(F^2-B^2)({\cal U}^2(p_2)+w^2(p_2))+
\nonumber\\
&&(2F^2+B^2)w(p_2)(w(p_2)-2\sqrt 2 Re {\cal U}(p_2))(3cos^2\theta_2 -1)+
\\
&&9B^2 w(p_2)(w(p_2)-2\sqrt 2 Re {\cal U}(p_2))sin^2\theta_2cos2\phi_2-
\nonumber\\
&&54(F^2-B^2)w^2(p_2)sin^2\theta_2cos^2\theta_2 sin^2\phi_2\}.
\nonumber
\end{eqnarray}
Note, that only squared nucleon- nucleon spin- flip amplitudes
$B^2$ and $F^2$ are in expression for the tensor analyzing power $C_{0,yy}$
and differential cross section. However, the spin correlation
due to vector polarization of deuteron and beam proton
 contains the interference
terms of this amplitudes
\begin{eqnarray}
C_{y,y}\cdot C_0&=&-{2\over {4 \pi}}\left( \frac{m_N+E_p}{2E_p}
\right) ^2
 Re(FB^*)\{2{\cal U}^2(p_2)-w^2(p_2)-
\nonumber\\
&&w(p_2)(w(p_2)+\sqrt 2 Re{\cal U}(p_2))(1-3sin^2\theta _2sin^2\phi_2)\}+
\\
&&12 w(p_2)Im(FB^*)Im{\cal U}(p_2)(cos^2\theta_2-sin^2\theta_2cos^2\phi_2)
\nonumber
\end{eqnarray} 
It is interesting, that there is the term proportional to the imaginary part
of ${\cal U}(p_2)$. It has a non-zero value only in case when FSI
is taken into account. The analogous result we have obtained for the
vector-tensor spin correlation
\begin{eqnarray}
\label{cyxz}
C_{y,xz}\cdot C_0&=&-{3\over {4 \pi}}\left( \frac{m_N+E_p}{2E_p}
\right) ^2
Im(FB^*)\{2{\cal U}^2(p_2)-w^2(p_2)-
\nonumber\\
&&w(p_2)(w(p_2)+\sqrt 2 Re{\cal U}(p_2))(1-3sin^2\theta _2sin^2\phi_2)+
\\
&&18w^2(p_2)sin^2\theta_2cos^2\theta_2cos^2\phi_2\}-
\nonumber\\
&&3\sqrt 2w(p_2) Re(FB^*)Im{\cal U}(p_2)(cos^2\theta_2-sin^2\theta_2cos^2\phi_2)
\nonumber
\end{eqnarray}
In order to evaluate these observables we consider kinematics, when
one of the slow protons is emitted along the beam direction as well as
neutron, i.e. $\theta_2=0^0$. Under such conditions Eqs. 
(\ref {c0})-(\ref {cyxz}) are significantly simplified. 
\begin{eqnarray}
\label{obs}
C_0&=&{1\over {2 \pi}}\left( \frac{m_N+E_p}{2E_p}
\right) ^2\{ (2B^2+F^2)({\cal U}^2(p_2)+w^2(p_2))+
\nonumber\\
&&(F^2-B^2)w(p_2)(w(p_2)-2\sqrt 2 Re{\cal U}(p_2))\},
\nonumber\\
\nonumber\\
C_{0,yy}\cdot C_0&=&{1\over {4 \pi}}\left( \frac{m_N+E_p}{2E_p}
\right) ^2
\{2(F^2-B^2)({\cal U}^2(p_2)+w^2(p_2))+
\\
&&(2F^2+B^2)w(p_2)(w(p_2)-2\sqrt 2 Re {\cal U}(p_2))\}
\nonumber\\
\nonumber\\
C_{y,y}\cdot C_0&=&-{2 \over {4 \pi}}\left( \frac{m_N+E_p}{2E_p}
\right) ^2
\{Re(FB^*)[2{\cal U}^2(p_2)-2w^2(p_2)-
\sqrt 2 Re{\cal U}(p_2)w(p_2)]-
\nonumber\\
&&3\sqrt 2 Im(FB^*)Im{\cal U}(p_2)w(p_2)\}
\nonumber\\
\nonumber\\
C_{y,xz}\cdot C_0&=&-{3 \over {4 \pi}}\left( \frac{m_N+E_p}{2E_p}
\right) ^2
\{Im(FB^*)[2{\cal U}^2(p_2)-2w^2(p_2)-
\sqrt 2 Re{\cal U}(p_2)w(p_2)]+
\nonumber\\
&&3\sqrt 2
Re(FB^*)Im{\cal U}(p_2)w(p_2)\}
\nonumber
\end{eqnarray}

The differential cross section and three polarization
observables are presented in  figs.(1-4). 
The Love and Franey parametrization with a set of parameters
obtained by fitting of the modern phase shift data SP00 \cite {ar, said}
has been employed for NN-amplitude.
The full lines correspond
to calculations with taking into account  both the FSI in the (pp)-pair
and S- and D- waves in the deuteron. The results obtained in
the PWIA are shown by the dashed lines. The dash-dotted lines 
in the figs.(2-4)
for polarization observables  are the full calculation
results without D-wave in the DWF. In fig.1 the full and dash-dotted
lines are undistinguished. All calculations were carried out with
Paris NN-potential \cite {NN} and Paris DWF \cite {Par}.

One can see, the FSI contribution to the differential cross section (fig.1)
is significant even at the very small proton momentum, while for the 
polarization observables the difference between PWIA and PWIA+FSI
is visible only for $p_2 \ge 10-15 $ MeV/c.  However, with increase
of the proton momentum up to 50 MeV/c the importance of the FSI
corrections to the PWIA also increases.

Note, the absolute value of the tensor analyzing power $C_{0,yy}$ (fig.2)
in the momentum interval of interest is near zero. In order to
understand the source of that, we disregard the D-wave in the DWF.
Then, as a consequence from Eq.(\ref {obs}),  the polarization
observables are defined
by the ratio of the nucleon-nucleon charge exchange amplitudes only
\begin{eqnarray}
\label{w0}
C_{0,yy}&=&{1\over 2}\cdot \frac {F^2-B^2}{2B^2+F^2}
\nonumber\\
C_{y,y}&=&-2\cdot \frac {Re(FB^*)}{2B^2+F^2}
\\
C_{y,xz}&=&-3\cdot\frac {Im(FB^*)}{2B^2+F^2}
\nonumber
\end{eqnarray}
Thus, the nearness of the tensor analyzing power to zero  indicates that
the absolute values of the spin-flip NN-amplitudes approximately
equal each other, $|B|\approx |F|$.

The vector-tensor spin correlation
$C_{y,xz}$ (fig.4) has also very small value,
$|C_{y,xz}|\approx 0.06$.
The magnitude of this observable decreases
up to zero for $p_2\approx 50$ MeV/c, if the FSI corrections and
D-wave in the deuteron are taken into account, while it is almost
 constant in the PWIA and PWIA+FSI without D-wave.
One can see from Eqs.(\ref {obs}, \ref {w0}) for $C_{y,xz}$, the
reason of this behaviour  is connected with the small
value of the imaginary part of the nucleon-nucleon amplitudes
product, $Im (FB^*)$. In such a way, the great contribution
into $C_{y,xz}$ gives the term proportional to $Re (FB^*)$,
which defined by D-wave and imaginary part of the generalized
function ${\cal U}(p_2)$. Note, that $Im {\cal U}(p_2)\ne 0$, if
FSI taken into account.

The other situation is for the vector-vector spin correlation 
$C_{y,y}$ (fig.3). The term
proportional to $Re(FB^*)$ gives also
a considerable contribution in this observable , but it is multiplied
on the ${\cal U}^2(p_2)$.
The magnitude of $C_{y,y}$ is close to the theoretical limit -2/3,
that confirms to the
conclusion about approximate equality of the nucleon-nucleon
amplitudes, $|B|$ and $|F|$. Besides, this allows to conclude, that the relative
phase between these amplitudes is close to zero.
It is seen from Eq.(\ref {w0}), where D-wave
was neglected.

Since all the considered observables are defined by 
the elementary nucleon-nucleon
amplitudes mostly, it is interesting to compare their behaviour
for different NN-parametrization.
In figs.(5-8) we present the same observables as in figs.(1-4) for
two sets of parameters. The full line corresponds to the parameterization
based on the modern shift analysis SP00 \cite {ar, said}.
The dashed line is obtained using a set of parameters for NN-amplitude
 from \cite {LF}.
One can see, the difference for the differential cross section (fig.5) is 
about 1.5-2 times. The absolute value for the tensor analyzing power
$C_{0,yy}$ (fig.6) with new parametrization is about 2.5 times smaller
than that with parametrization \cite {LF}. The opposite situation is
for the vector-vector spin correlation $C_{y,y}$ (fig.7), where  new predictions
is 1.5 times lager in comparison with old parametrization of the NN-amplitude
\cite {LF}. The predictions of these two parameterizations for the vector-
tensor spin correlation $C_{y,xz}$ (fig.8) are even in opposite sign.
Nevertheless, the qualitative behaviour  of the curves
in figs. (6-8) for different sets of parameters is similar.

\section {Conclusion}

We have studied the deuteron -proton charge exchange reaction at
1 GeV energy in special kinematics, $\vec q \approx 0$. The influence
of the  D-wave in the deuteron and
FSI between two slow protons has been considered. 
It was shown, that D-wave and FSI
effects  are negligible for the polarization observables at the
proton momentum up to 10-15 MeV/c. As the result, in this region the
polarization observables are defined by the ratio of the nucleon-
nucleon charge exchange amplitudes only.
However, it  should not be ignored, that
 importance of the D-wave and ,
especially, FSI into polarization observables increases at 
$p_2 \ge 15$ MeV/c. In such a way, we conclude, that
 ratio of the nucleon-
nucleon charge exchange amplitudes and phase shift between their can
be extracted from experimental data rather simple, if the experimental
conditions and technical setup possibilities
allow to work in this small momentum interval. In opposite case,
this procedure is more complicated and  model dependent.
 It should be remembered,
that FSI contribution into the differential cross section is very
significant in comparison with PWIA predictions even at very small
proton momentum. This fact does not enables us to get the absolute
value of the nucleon-nucleon spin flip amplitudes without considering
the FSI corrections.

\vspace{2cm}
We are grateful to V.V.Glagolev, M.S.Nioradze and A.Kacharava for
inspiration of interest to this problem.
The authors are thankful to V.P. Ladygin for fruitful discussions. 


\newpage
\begin{center}
{\large\bf Figure captions}
\end{center}
\vspace{1cm}

{\bf Fig.1} The differential cross section at $\vec q=0$ as 
a function of one of the slow proton momentum. The dashed and full line correspond
to the PWIA and PWIA+FSI, respectively.

{\bf Fig.2} The tensor analyzing power $C_{yy}$ vs. $p_2$. The dashed line corresponds 
to  PWIA; dash-dotted and full lines are PWIA+FSI without D-component in the DWF
and with it, respectively.

{\bf Fig.3} The spin-correlation $C_{y,y}$ due to the vector polarization of 
the deuteron. The curves are the same as in fig.2.

{\bf Fig.4} The spin-correlation $C_{y,xz}$ due to the tensor polarization of
the deuteron. The curves are the same as in fig.2.

{\bf Fig.5} The differential cross section at $\vec q=0$ as 
a function of $p_2$. The dashed and full line correspond to the full calculation
with a set of NN-amplitude  parameters  taken from \cite {LF} and fit of SP00 data
 \cite {said}.
 
{\bf Fig.6} The tensor analyzing power $C_{yy}$ vs. $p_2$.
The curves are the same as in fig.5.

{\bf Fig.7} The spin-correlation $C_{y,y}$ due to the vector polarization of 
the deuteron. The curves are the same as in fig.5.

{\bf Fig.8} The spin-correlation $C_{y,xz}$ due to the tensor polarization of
the deuteron. The curves are the same as in fig.5.

\newpage
\vspace*{20cm}

\begin{figure}[h]
\includegraphics{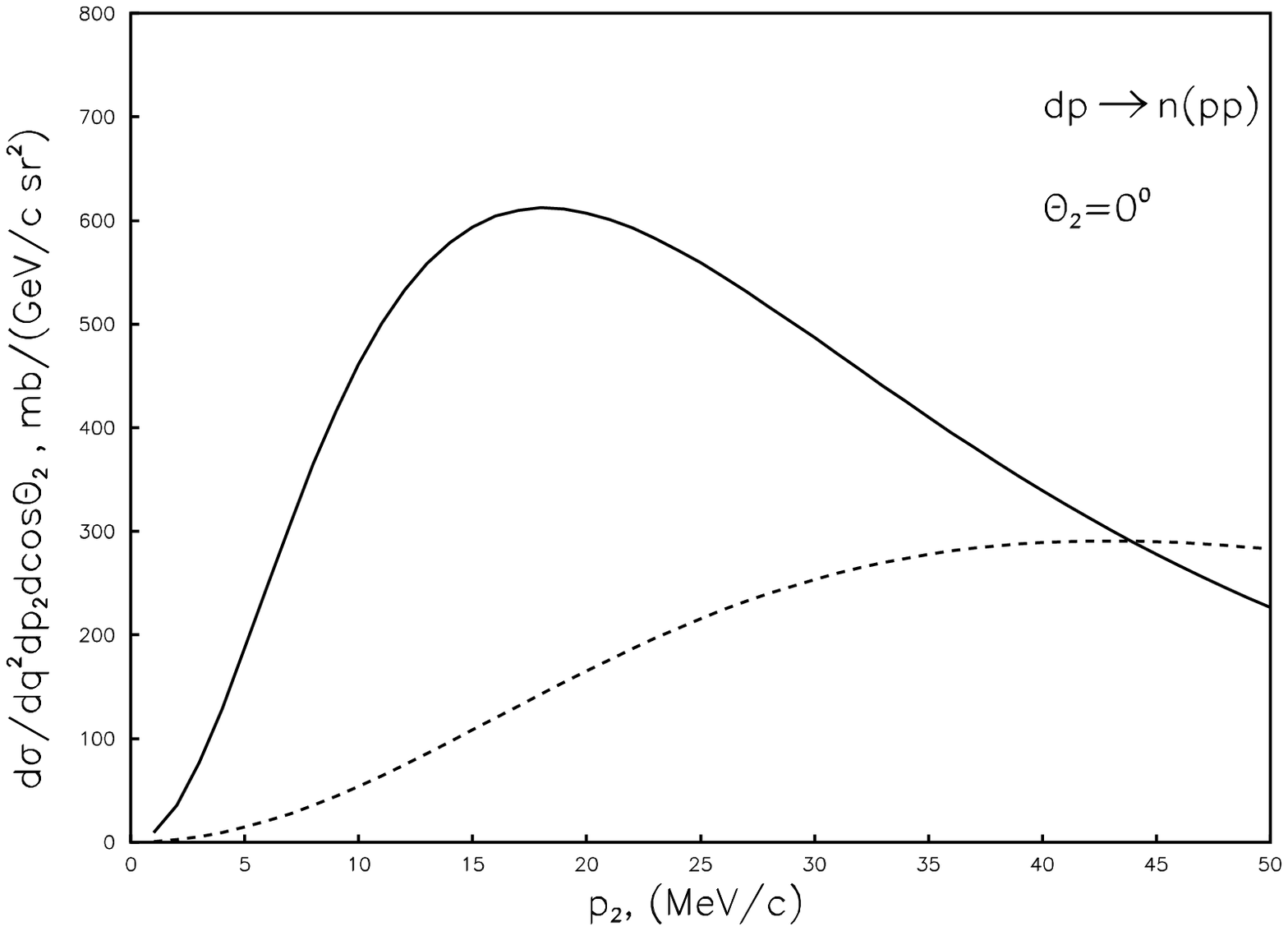}
\end{figure}
\vspace*{-5cm}
{\bf Fig.1}

\newpage
\vspace*{20cm}

\begin{figure}[h]
\includegraphics{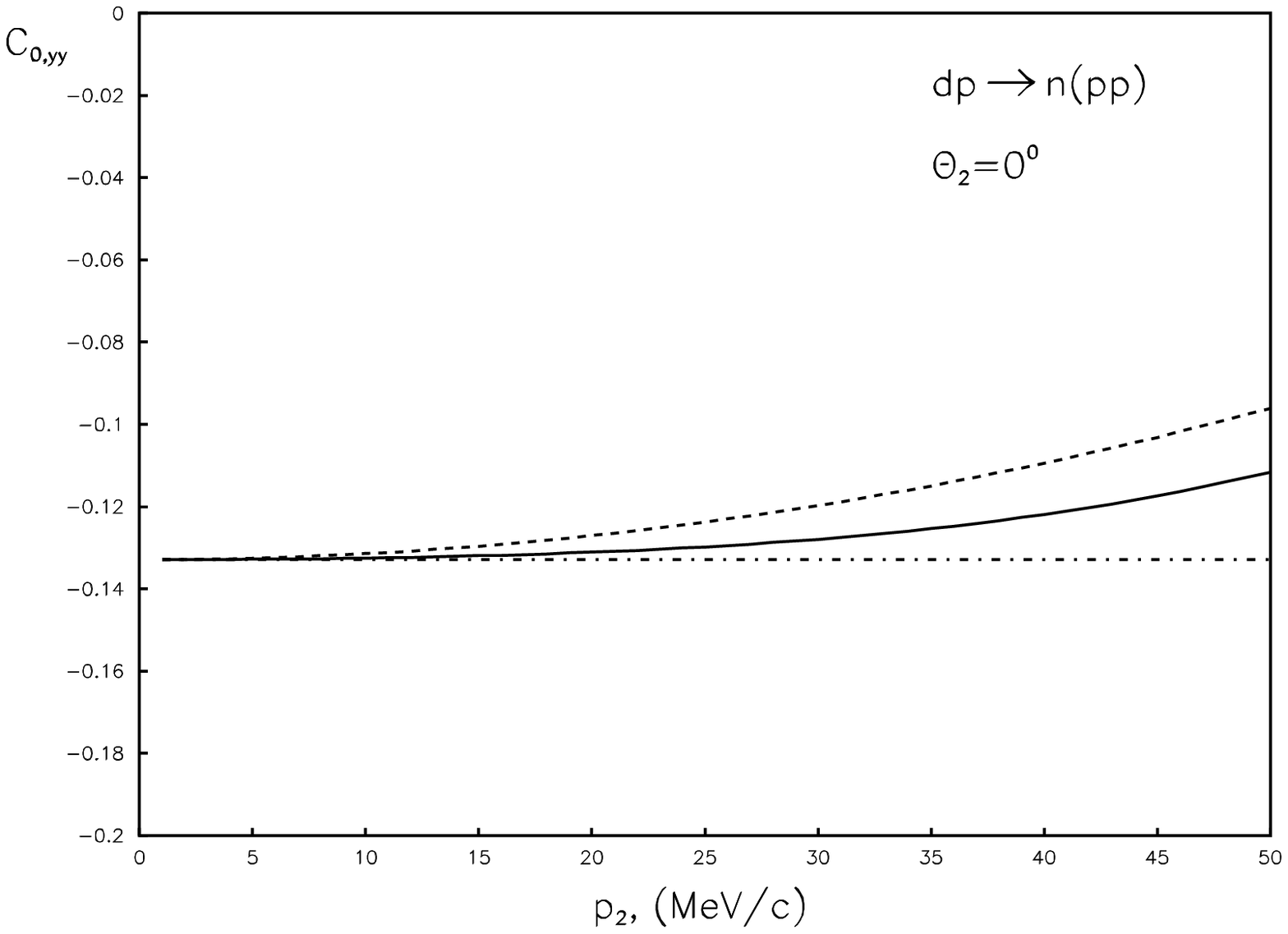}
\end{figure}
\vspace*{-5cm}
{\bf Fig.2}
\newpage
\vspace*{20cm}

\begin{figure}[h]
\includegraphics{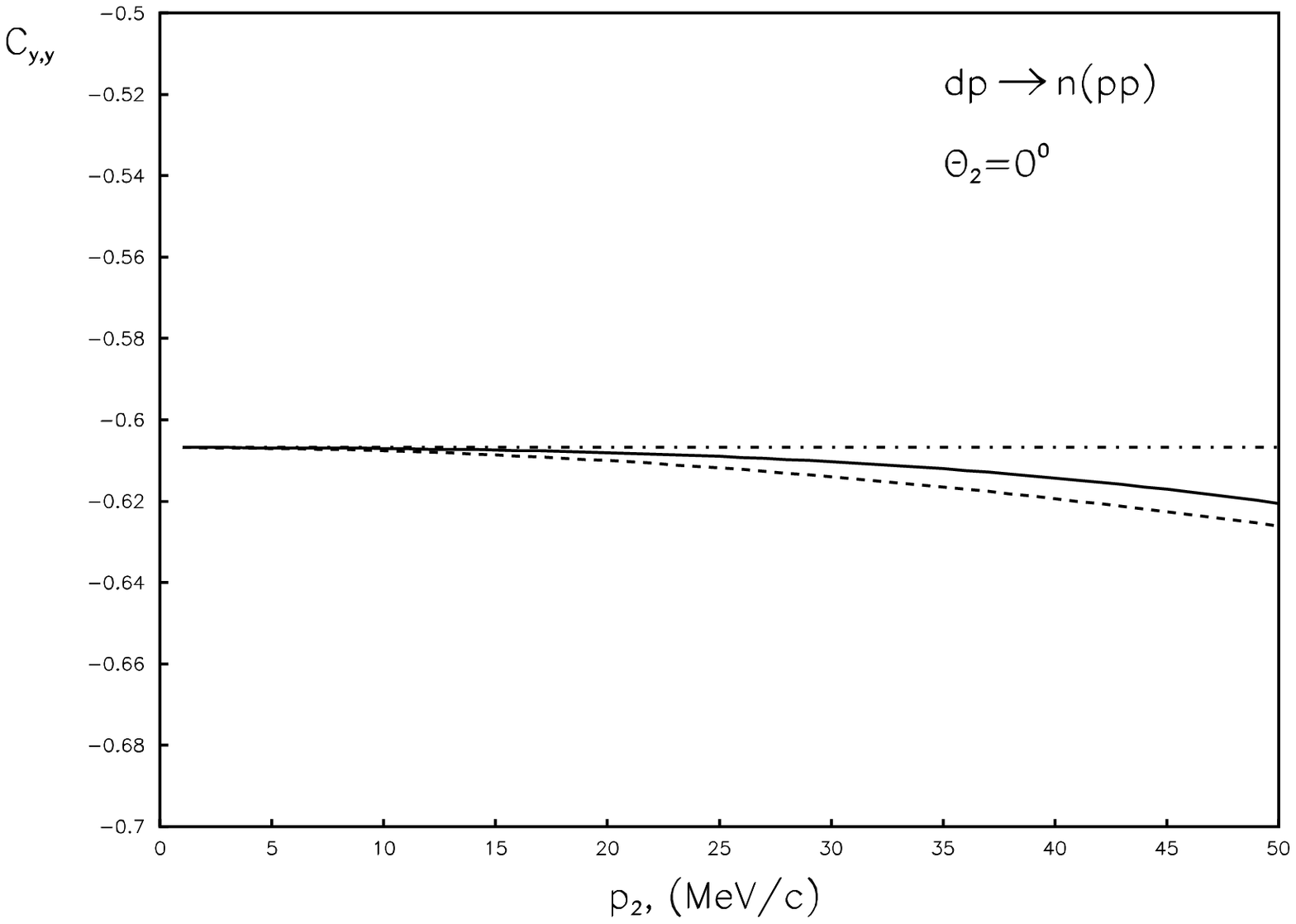}
\end{figure}
\vspace*{-5cm}
{\bf Fig.3}
\newpage
\vspace*{20cm}

\begin{figure}[h]
\includegraphics{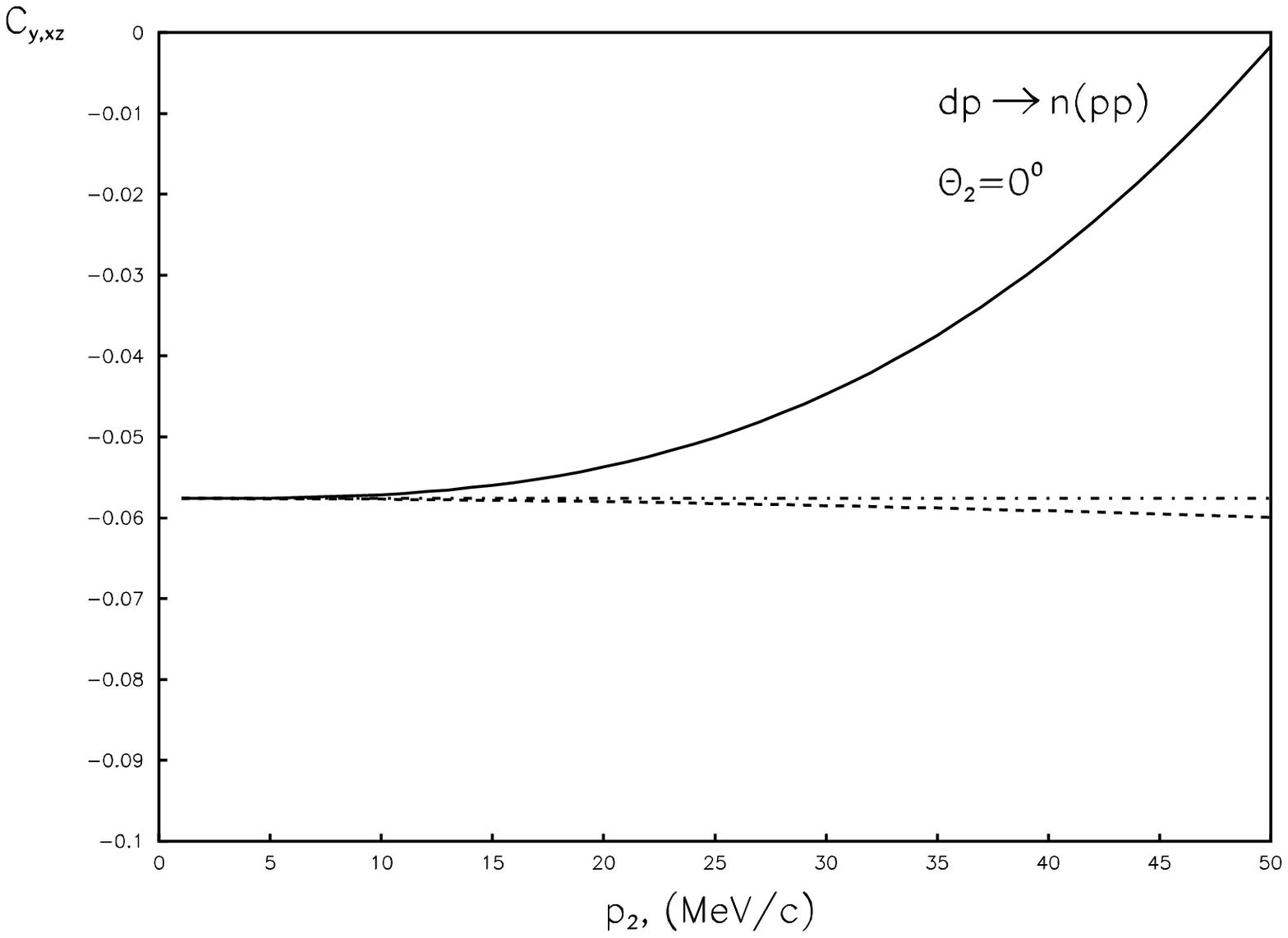}
\end{figure}
\vspace*{-5cm}
{\bf Fig.4}
\newpage
\vspace*{20cm}

\begin{figure}[h]
\includegraphics{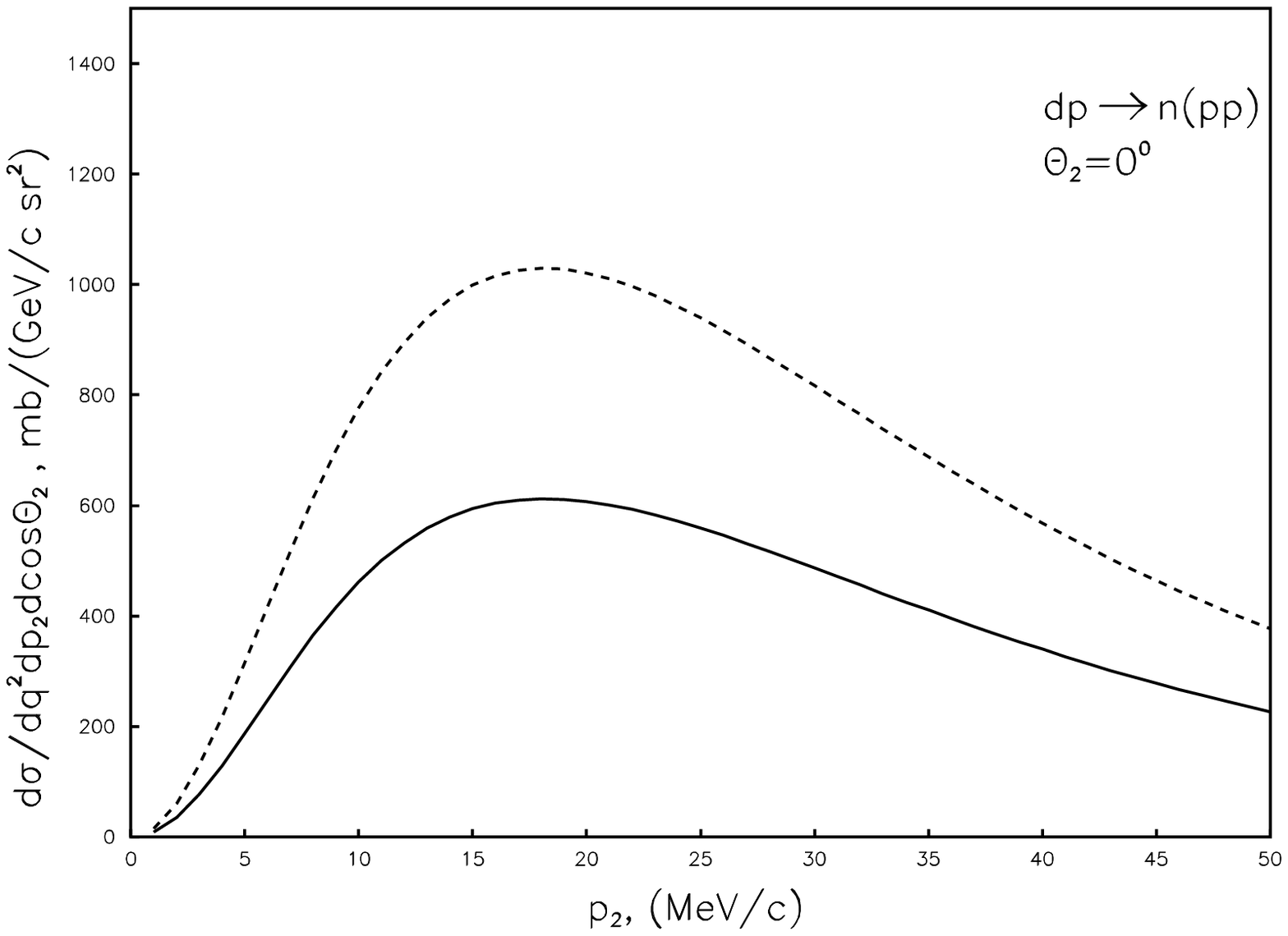}
\end{figure}
\vspace*{-5cm}
{\bf Fig.5}
\newpage
\vspace*{20cm}

\begin{figure}[h]
\includegraphics{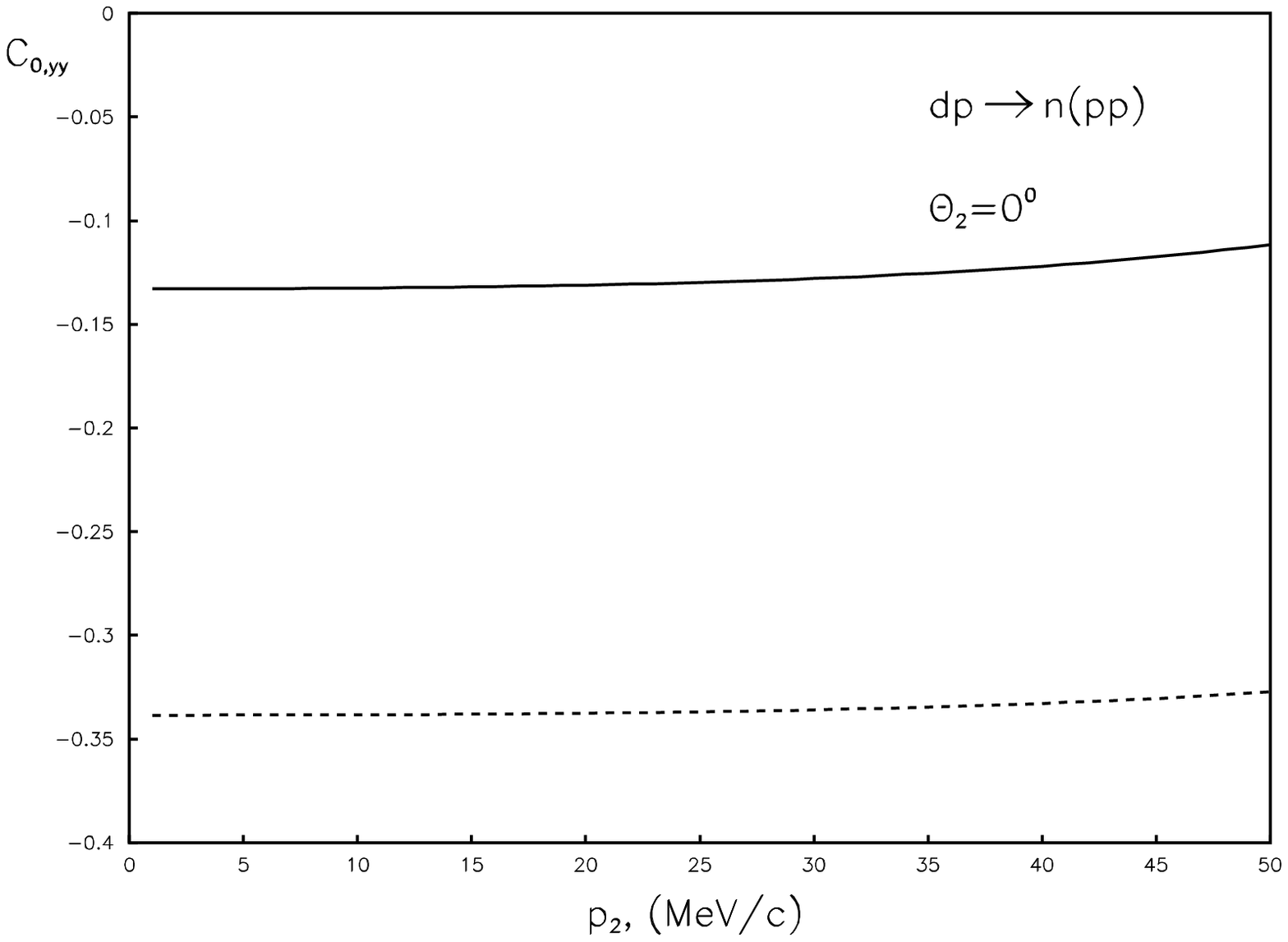}
\end{figure}
\vspace*{-5cm}
{\bf Fig.6}
\newpage
\vspace*{20cm}

\begin{figure}[h]
\includegraphics{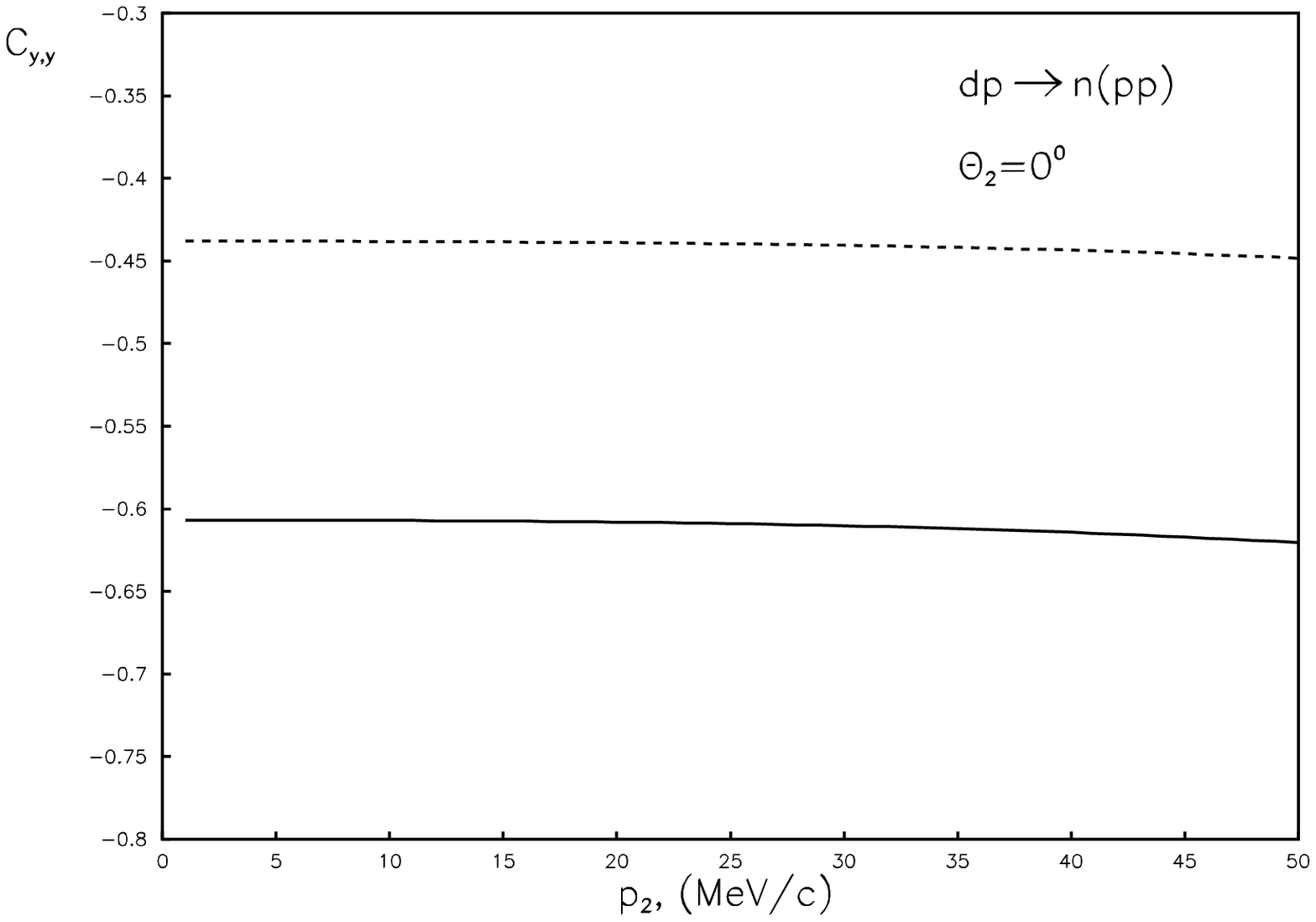}
\end{figure}
\vspace*{-5cm}
{\bf Fig.7}
\newpage
\vspace*{20cm}

\begin{figure}[h]
\includegraphics{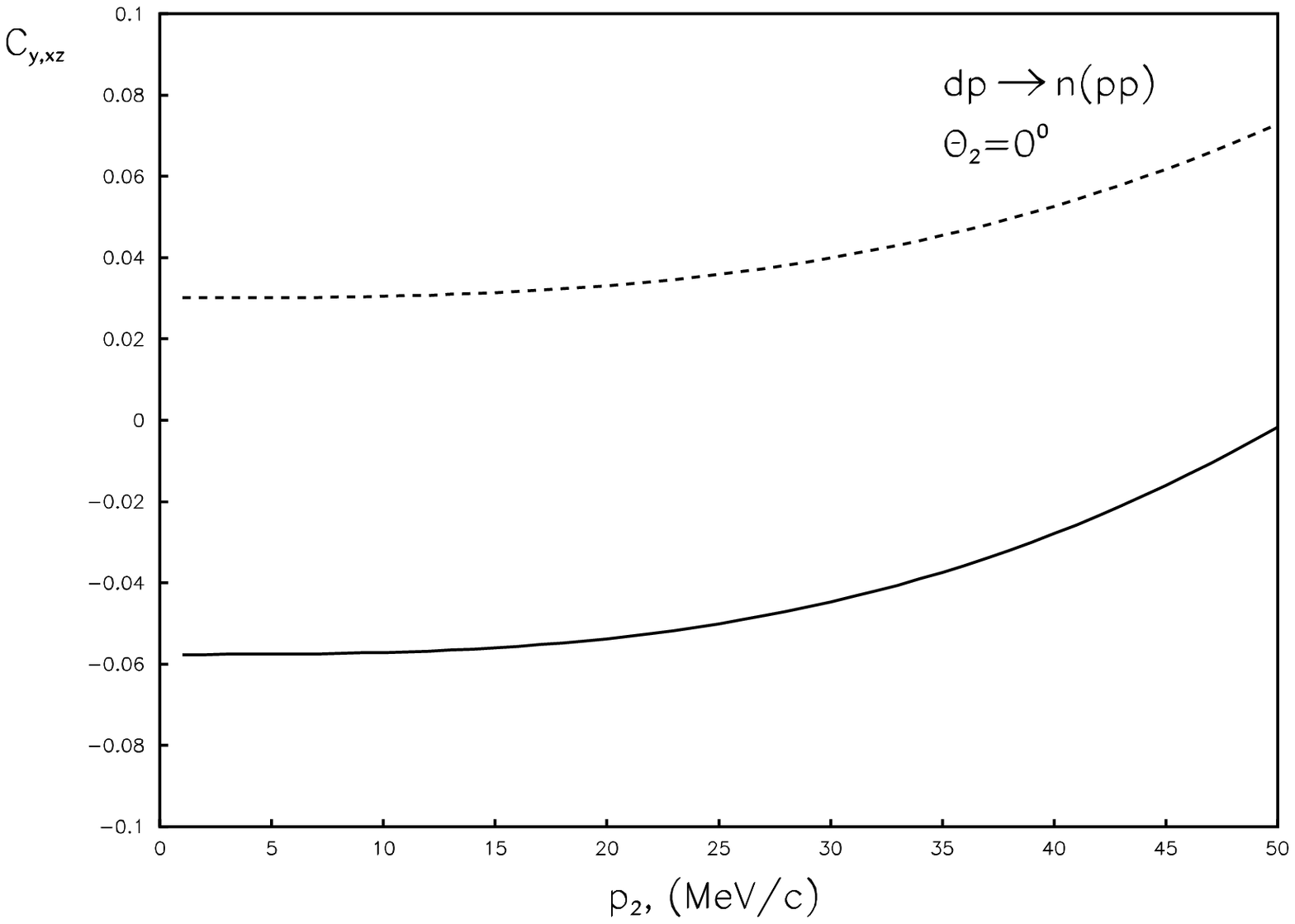}
\end{figure}
\vspace*{-5cm}
{\bf Fig.8}

\end{document}